\documentclass[a4paper,11pt]{article}
\usepackage{pos}
\usepackage{subcaption}
\usepackage{enumitem}
\newcommand{\Xmax}{\ensuremath{\mathrm{X}_{\mathrm{max}}} }
\newcommand{\Amax}{\ensuremath{\mathcal{A}_{\mathrm{max}}} }
\newcommand{\Smax}{\ensuremath{\mathcal{S}_{\mathrm{max}}} }
\usepackage[cal=cm]{mathalfa}

\newlength{\bibitemsep}
\newlength{\bibparskip}
\let\oldthebibliography\thebibliography
\renewcommand\thebibliography[1]{\oldthebibliography{#1}\setlength{\parskip}{\bibitemsep}\setlength{\itemsep}{\bibparskip}%
}

\title{Exploring the sensitivity of in-ice radio detectors to cosmic ray mass composition}

\author[a]{Nathaniel Alden}
\author[a]{Marissa Boucher}
\author*[a]{Cosmin Deaconu}
\author[a]{Abigail Vieregg}
\author[a,b]{Philipp Windischhofer}

\affiliation[a]{Dept. of Physics, Dept. of Astronomy and Astrophysics, Enrico Fermi Inst., Kavli Inst. for Cosmological Physics, University of Chicago,\\
  Chicago, IL 60637, USA}

\affiliation[b]{Deutsches Elektronen-Synchrotron DESY,\\
Platanenallee 6, 15738, Zeuthen, Germany}

\emailAdd{nalden@uchicago.edu}
\emailAdd{mboucher28@uchicago.edu}
\emailAdd{cozzyd@kicp.uchicago.edu}
\emailAdd{avieregg@kicp.uchicago.edu}
\emailAdd{philipp.windischhofer@desy.de}

\abstract{In-ice radio detectors have been developed primarily for the detection of high-energy neutrinos via the Askaryan effect, but have recently been shown to also be sensitive to cosmic ray air showers impacting the ice sheet.
Using CORSIKA 8 to simulate impacting air showers, we find that the lateral width of the in-ice cascade is sensitive to the atomic mass of the primary cosmic ray.
The width of the shower core is accessible through the shape of the Askaryan radio spectrum, while the amplitude encodes the total in-ice charge excess.
Using these two observables, we show that in-ice radio detectors can be used to measure cosmic ray mass composition in the region above $10^{17}$\,eV if the in-ice receivers are placed to adequately sample the Askaryan emission pattern.
We also show that the in-ice core width is highly, but not exactly, correlated with the depth of maximum particle count (\Xmax), the most commonly used observable for reconstructing cosmic ray mass composition.
}

\FullConference{11th International Workshop on Acoustic and Radio EeV Neutrino Detection Activities (ARENA2026)\\
8-11 June 2026\\
Karlsruhe, Germany\\}

\begin{document}
\maketitle

\subsection*{Introduction}
\noindent
Cosmic rays are the highest-energy particles ever observed and are a valuable probe of the highest-energy astrophysical processes.
Breaks in the energy spectrum likely correspond to a change in the acceleration mechanism,
while measurements of the mass composition as a function of energy provide even more detail about astrophysical accelerators \cite{Kotera_CR_overview, UHECR_aboveknee_review, mass_comp_below_ankle_review}.
It remains an open question which sources are responsible for the cosmic ray flux between the `second knee' spectral feature at $10^{17.5}$\,eV \cite{hires_ankle, yakutsk_second_knee, tunka_second_knee, hires_second_knee, Kascade_grande_second_knee, IceTop_second_knee, second_knee_review}, thought to be the cutoff of Galactic cosmic rays, and the `ankle' feature at $10^{18.6}$\,eV \cite{hires_ankle, auger_ankle}, presumed to indicate the dominance of extragalactic cosmic rays \cite{mass_comp_below_ankle_review, Unger_transition}.

Cosmic ray astrophysics is complicated by intervening magnetic fields affecting the trajectory of the charged cosmic rays, impeding source reconstruction. Furthermore, the highest-energy cosmic rays cannot propagate over cosmological distances due to the GZK mechanism~\cite{GZK_G,GZK_ZK}.
This has in part motivated the search for ultrahigh-energy cosmic neutrinos, as neutrinos are uncharged and weakly interacting---they therefore both point back to their sources and can propagate through the entire observable universe \cite{uhecr_neutrinos}.
Optical neutrino detectors such as IceCube \cite{IceCube_instrument} and KM3NeT \cite{KM3NeT_instrument} have observed high-energy astrophysical neutrinos \cite{KM3NeT_highenergyneutrino} and identified a small number of potential sources \cite{icecube_NGC1068, icecube_TXS}.
At higher energies, even larger detectors are necessary to resolve the extremely low flux, and the radio detection method has emerged as a promising experimental technique \cite{ZHS92}.
The Askaryan Radio Array (ARA)~\cite{ARA_instrument_paper} and the Radio Neutrino Observatory Greenland (RNO-G)~\cite{RNO-G_instrument_paper} are two experiments employing antennas placed 100\,m--200\,m into the polar ice sheets.
They aim to make use of the Askaryan mechanism \cite{askaryan_1, askaryan_2} and the km-scale radio attenuation length in ice~\cite{atten_length_radio} to detect ultrahigh-energy neutrinos. 
Additionally, the proposed IceCube-Gen2 experiment includes a radio component in addition to an expanded optical component \cite{icecubegen2_whitepaper}.

Recently, it has been demonstrated that in-ice radio detectors such as ARA and RNO-G can also detect cosmic rays via the Askaryan emission produced when the core of a cosmic ray air shower impacts the polar plateau, at an altitude of about 3000\,m above sea level~\cite{ARA_CRs}. The expected rate of cosmic ray observations using this technique is estimated to be $\mathcal{O}(10-100)$ per station-year~\cite{ARA_CRs, coleman2024iniceaskaryanemissionair}.
For reference, RNO-G currently has 12 antenna stations operating near Summit Station, Greenland, while ARA has already collected $\sim24$ station-years of data at South Pole and continues to operate a smaller number of stations.
The primary energy threshold for the detection of cosmic rays via their Askaryan emission is approximately $10^{17}$\,eV \cite{ARA_CRs, coleman2024iniceaskaryanemissionair}, somewhat lower than at large cosmic ray observatories such as the Telescope Array \cite{TA_fluorescence} and the Pierre Auger Observatory \cite{Auger_fluorescence} due to the high altitude ($\sim3\,\mathrm{km}$) at both South Pole and Summit Station.
We therefore expect a significant flux of cosmic rays in an energy range which is conveniently situated to better understand the transition between Galactic and extragalactic cosmic rays.
However, the Askaryan signal is primarily generated by the in-ice shower core \cite{Seckel_CR_cores, Simon_Krijn_CR_cores, ARA_CRs} and therefore affected by the stochastic shower development in the atmosphere.

Whether it is possible to use the in-ice Askaryan signature to reconstruct the primary energy and atomic mass of the cosmic ray is therefore both a natural question and an open problem.
In this contribution, we show that the preliminary answer is affirmative given the capabilities of currently operating detectors like ARA and RNO-G.

\subsection*{Mass sensitivity of the shower core}
\noindent
When a sufficiently high-energy cosmic ray reaches the ice surface, only the particles within a radius of $\sim10$\,cm around the shower axis are responsible for the in-ice radio emission. 
The energy in this core is primarily carried by high-energy photons which seed the in-ice cascade and create a net charge excess through the Askaryan effect \cite{ARA_CRs}.
In contrast, the in-air charge distribution, responsible for radio emission via the geomagnetic effect, has a much larger lateral extent than the in-ice charge excess~\cite{ARA_CRs}.
The in-air lateral charge profile is an established discriminator of the primary mass~\cite{original_electron_lateral, yakutsk_lateral, KASCADE-GRANDE_experiment}. 
The inner core width has also been used by the ARGO experiment to resolve primary mass, but only at lower energies where the flux is high enough to allow observation with dense particle detectors~\cite{argo_lateral}.

\begin{figure}[t]
\centering
\begin{subfigure}{.5\columnwidth}
  \centering
  \includegraphics[width=.9\columnwidth]{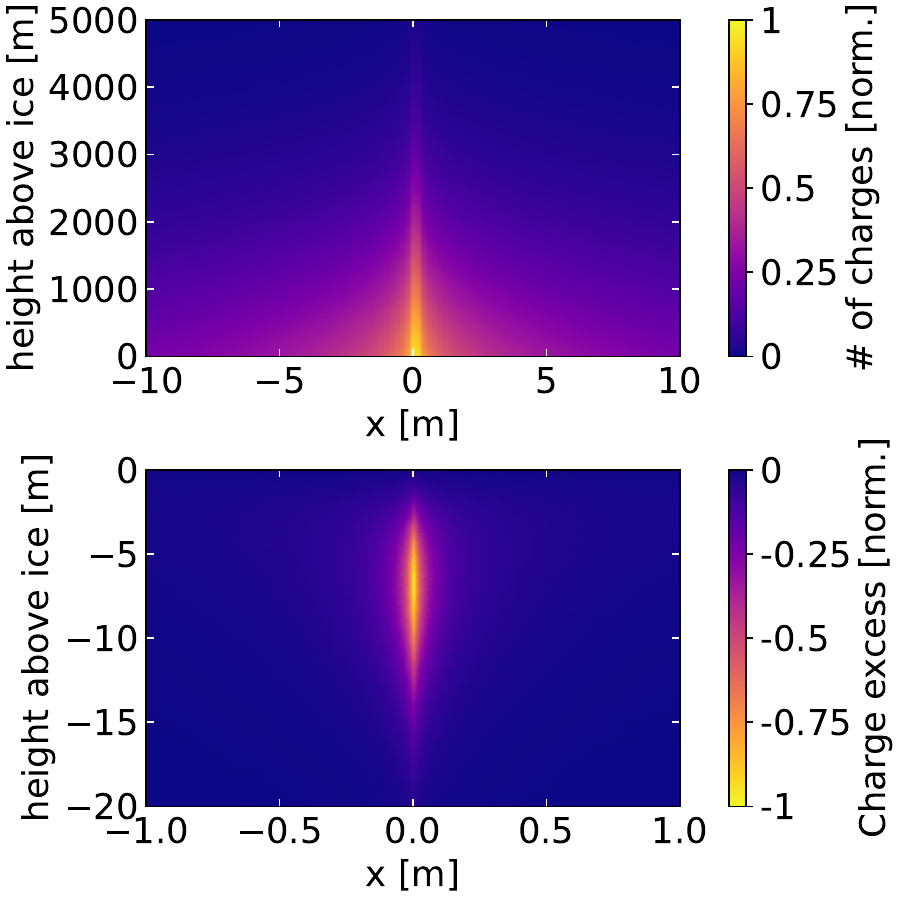}
  \label{fig:sub1}
\end{subfigure}%
\begin{subfigure}{.5\columnwidth}
  \centering
  \includegraphics[width=.9\columnwidth]{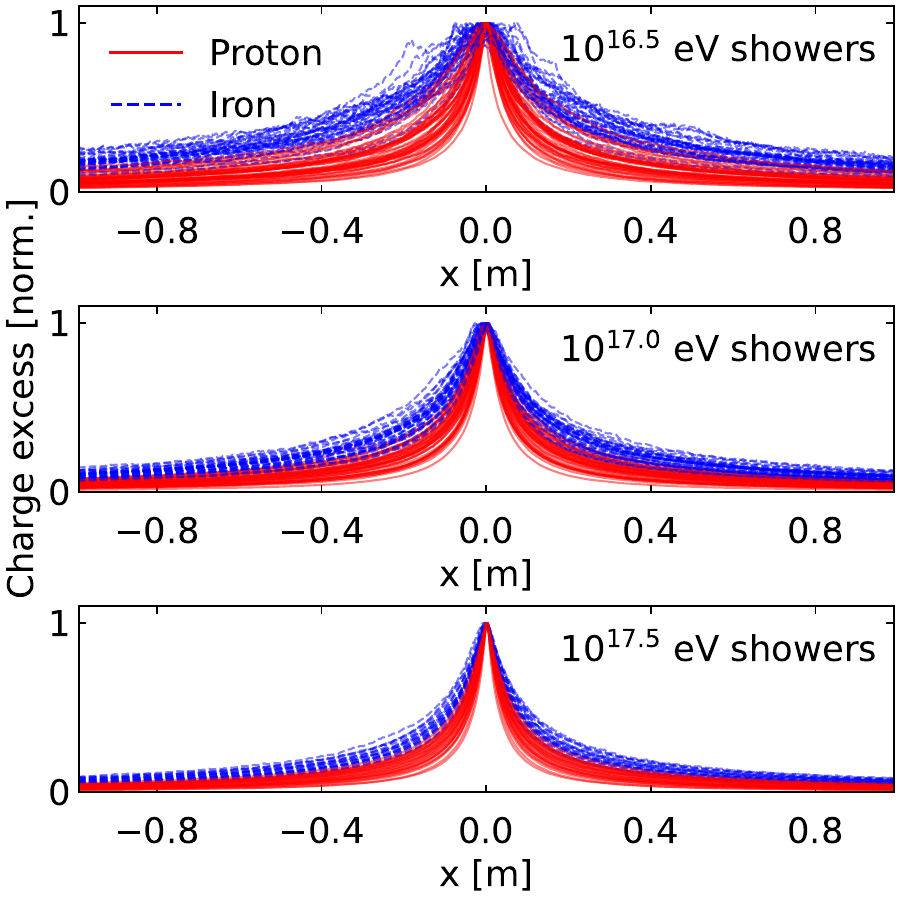}
  \label{fig:sub2}
\end{subfigure}
\caption{Left: the shape of the in-air charge profile (top) and the in-ice charge excess profile (bottom) of a shower generated using CORSIKA 8 \cite{corsika8}.
These correspond to the dominant source of radio emission in each medium. 
Note the different lateral scales: the in-ice charge excess is much more collimated than the in-air charge profile.
Right: Overlaid in-ice charge profiles, obtained by marginalizing over the longitudinal shower development. 
Showers initiated by proton (red) and iron (blue) primaries are shown at three different primary energies.}
\label{fig:charge_excess}
\end{figure}

Here we use the CORSIKA 8 software \cite{corsika8} to simulate the microscopic shower development and to calculate the radio emission using the ZHS formalism \cite{ZHS92}.
We use the default particle thinning setting for increased simulation speed.
The in-air charge distribution and the in-ice charge excess distribution for one such simulated shower are shown in Fig.~\ref{fig:charge_excess} (left), where the different x axes highlight the difference in lateral scale.
The right side of Fig.~\ref{fig:charge_excess} shows the lateral profile of just the in-ice charge excess.
We observe that the in-ice lateral core width is sensitive to primary mass, as previously investigated in Refs.~\cite{Seckel_CR_cores, Simon_Krijn_CR_cores}, and we also observe that the separation between proton and iron persists with increasing energy.
There is also an overall trend towards narrower cores with increasing energy, indicating the need for two independent observables to reconstruct primary energy and mass.
The need for two observables is shared by traditional air-shower detectors---for example, the \Xmax and total fluorescence light for fluorescence detectors~\cite{Auger_fluorescence, TA_fluorescence}, or the amplitude and shape of the particle footprint as measured by ground-based particle detectors~\cite{KASCADE-GRANDE_experiment, TA_surface_detector, Auger_watercherenkov, Auger_scintillators}.
\begin{figure}[t]
    \centering
  \includegraphics[width=0.75\columnwidth]{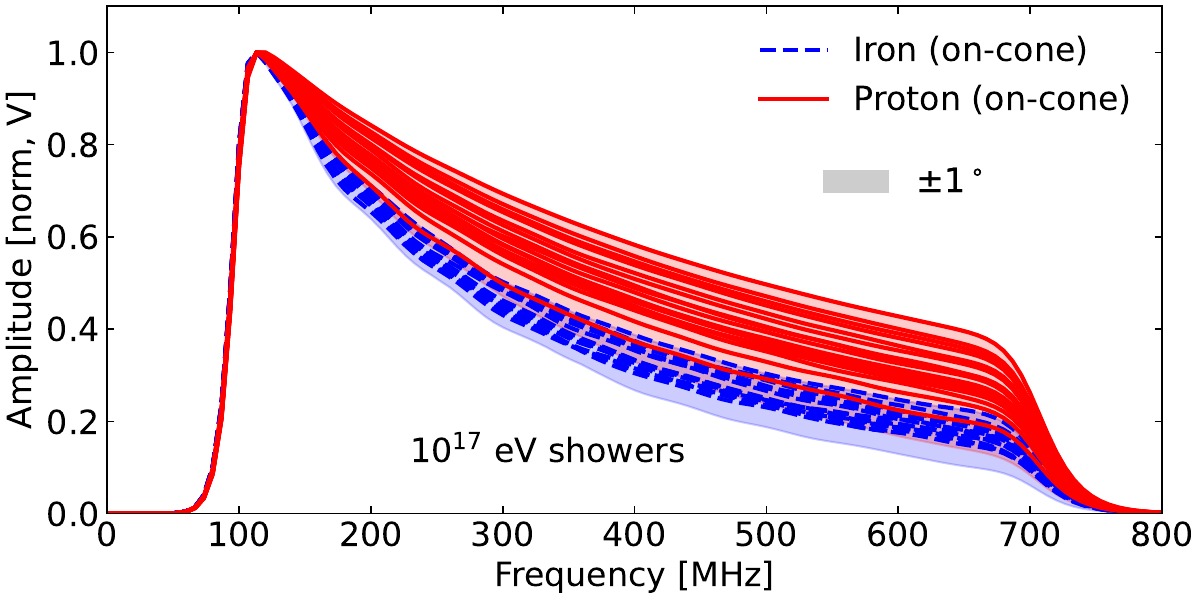}
  \caption{
The in-ice Askaryan radio spectrum for $10^{17}$\,eV proton and iron showers normalized to the peak spectral power, with the shaded region indicating a $\pm1^\circ$ range of the observation angle relative to the Cherenkov angle. To simulate a realistic antenna and signal chain, a $1/f$ antenna effective height and a 10th-order Butterworth bandpass filter from 100\,MHz to 700\,MHz were applied to the electric field.
  }
  \label{fig:spectra}
\end{figure}

\subsection*{Imprint on the radio spectrum}
\noindent
The Askaryan signal is characterized by a fast, coherent, and bipolar radio pulse.
In the frequency domain, the pulse power increases with frequency until a characteristic frequency where coherence begins to be lost.
This low-pass cutoff frequency depends on the observation direction relative to the Cherenkov angle, in addition to the lateral and longitudinal scales of the in-ice shower.
The spectral effect of each length scale has a distinct angular dependence.
In particular, when the shower is viewed at the Cherenkov angle, there is perfect longitudinal coherence and the only low-pass effect comes from the mass-dependent lateral size of the shower.
At a characteristic off-cone angle
given by $\delta\theta_\mathrm{crit} = |\theta_\mathrm{crit}-\theta_c| \approx R/L$ \cite{AMVZ06}, the spectral low-pass effect begins to be dominated by loss of coherence in the longitudinal direction, diluting the mass sensitivity.

Typical in-ice shower cores have $R \sim 10$\,cm and $L \sim 5$\,m, giving $\delta\theta_\mathrm{crit} \sim 1^\circ$. 
We therefore expect that, within $1^\circ$ of the Cherenkov cone, the shape of the Askaryan radio spectrum can be used to discriminate mass composition.
This observation is confirmed in Fig.~\ref{fig:spectra}, which shows a clear separation between proton and iron spectral shape for showers observed within $1^\circ$ of the Cherenkov cone.
As expected from the energy dependence of the core width, there is also an energy dependence in the shape of the radio spectrum: in general, higher energy showers have a narrower core and therefore more high-frequency content.

There is also a zenith dependence, since more-inclined showers have a longer path length in the atmosphere and, consequently, a more diffuse in-ice core.
However, it is possible to reconstruct cosmic ray zenith to degree-level accuracy using just a few surface components, such as scintillators or upward-facing antennas~\cite{sjoerd_thesis}.
Such surface antennas are already deployed at RNO-G \cite{RNO-G_instrument_paper} and are a planned component at IceCube-Gen2 radio \cite{icecubegen2_whitepaper}.
For the present study, we therefore assume that the zenith dependence can be controlled using a sparse surface detector
and focus on designing observables to reconstruct just the primary mass and energy from the in-ice Askaryan radio spectrum.

\subsection*{Defining experimentally robust observables}
\noindent
We have seen that the shape of the in-ice Askaryan radio spectrum provides sensitivity to primary mass, but also depends on energy.
To break this degeneracy, a natural choice is to introduce an amplitude observable, since higher-energy showers create a larger in-ice charge excess, and therefore a larger amplitude radio signal.
To maximize the visibility of the signal over flat-spectrum thermal noise from the surrounding ice, we make use of the spectral power integrated over wide frequency bands.
We use a low-frequency (LF) band (0-250\,MHz) and a high-frequency (HF) band (250-500\,MHz).
Then we define the amplitude observable $\Amax$,
\begin{equation}
    \mathcal{A}_\mathrm{max} = \mathrm{max}\left (\mathrm{power(LF)}\right),
\end{equation}
where the max is taken over all antennas observing an event, and the shape observable $\Smax$,
\begin{equation}
    \mathcal{S}_\mathrm{max} = \left.\frac{\mathrm{power(HF)}}{\mathrm{power(LF)}}\right|_\mathrm{argmax(power(LF))}.
\end{equation}
These observables are intended to be interpretable measures of the spectral amplitude and shape.
More complicated observables providing improved performance can likely be found.

We have also seen that the imprint of the lateral shape is strongest close to the Cherenkov cone.
Experiments with densely-spaced in-ice antennas are likely to always have at least one receiver placed very near the Cherenkov angle,
and are therefore expected to achieve better mass discrimination power.
It may also be possible to combine signals from all antennas and extrapolate to the Cherenkov angle, although we do not provide a prescription to do this here.
To demonstrate the mass discrimination capabilities achievable in  different situations, we consider the following three cases:
\begin{enumerate}[label=(\roman*)]
    \item The ideal case in which the detector always measures the true \Amax{}. This is representative of a densely-spaced antenna array in which one receiver observes the cascade under the Cherenkov angle.
    \item Assuming that the on-cone Askaryan signal can be observed at a viewing angle $\theta$ such that the off-cone angle $\theta - \theta_c$ follows a normal distribution with a standard deviation of $0.5^\circ$, representative of the case where an effort is made to extrapolate the position of the Cherenkov cone based on information from multiple antennas.
    \item Assuming that the Askaryan signal can be observed with $\theta - \theta_c$ distributed uniformly within $\pm2^\circ$ of the Cherenkov angle. This case is representative of an experiment with an RNO-G-like geometry in which the antenna with the largest observed signal amplitude is used for the mass analysis.
\end{enumerate}

\begin{figure}[t]
    \centering
  \includegraphics[width=0.75\columnwidth]{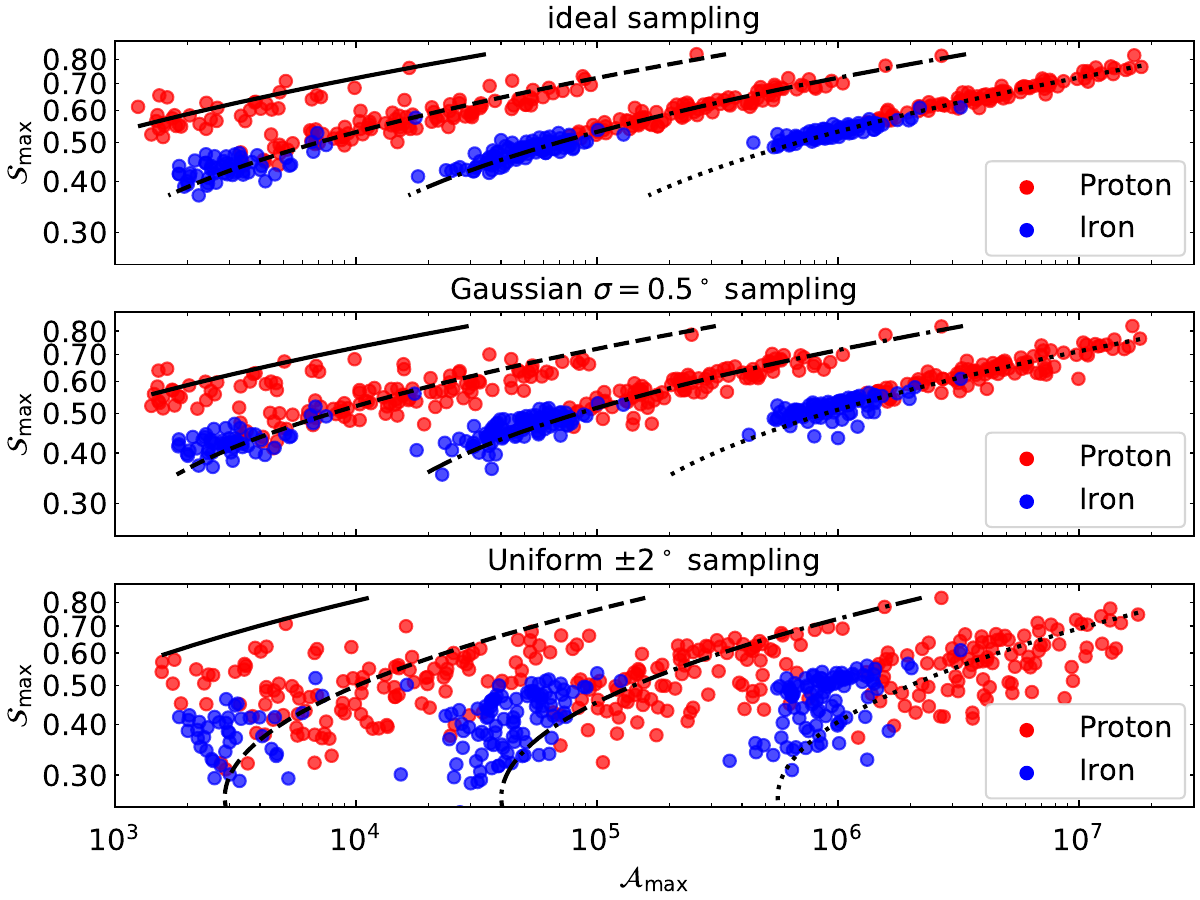}
  \caption{
The radio observables \Smax and \Amax, defined in the text, with 100 proton and iron showers simulated at each energy before applying an $\mathrm{SNR}=3$ cut. Black lines show constant-energy contours: from left, the lines are for $10^{16.5}$\,eV, $10^{17}$\,eV, $10^{17.5}$\,eV, and $10^{18}$\,eV showers, respectively. The three panels show the result of three different assumptions, also described in the text, on the experimental ability to identify the radio signal closest to the peak of the Cherenkov cone.
  }
  \label{fig:radio_obs}
\end{figure}

In Fig.~\ref{fig:radio_obs} we show the radio observables \Amax and \Smax under each of these experimental assumptions.
As the identification of the on-cone signal gets less precise, we observe two trends.
First, there is a bias towards smaller \Smax as more events miss the peak of the Cherenkov cone, where \Smax is largest.
Second, the intrinsic scatter increases.
Both of these effects would need to be accounted for using an accurate simulation of the detector.
The bias needs to be corrected for in order to give an unbiased estimate of the primary mass and energy, while the scatter must be accounted for when estimating the per-event resolution on each quantity.
Nevertheless, in all three cases we observe nonzero separation between the proton and iron distributions, even after acknowledging that the true energy distribution is continuous and not discrete.
In particular, case (ii) achieves mass separation that is qualitatively similar to the ideal case, turning this into an important experimental benchmark scenario.

\begin{figure}[t]
    \centering
  \includegraphics[width=0.75\columnwidth]{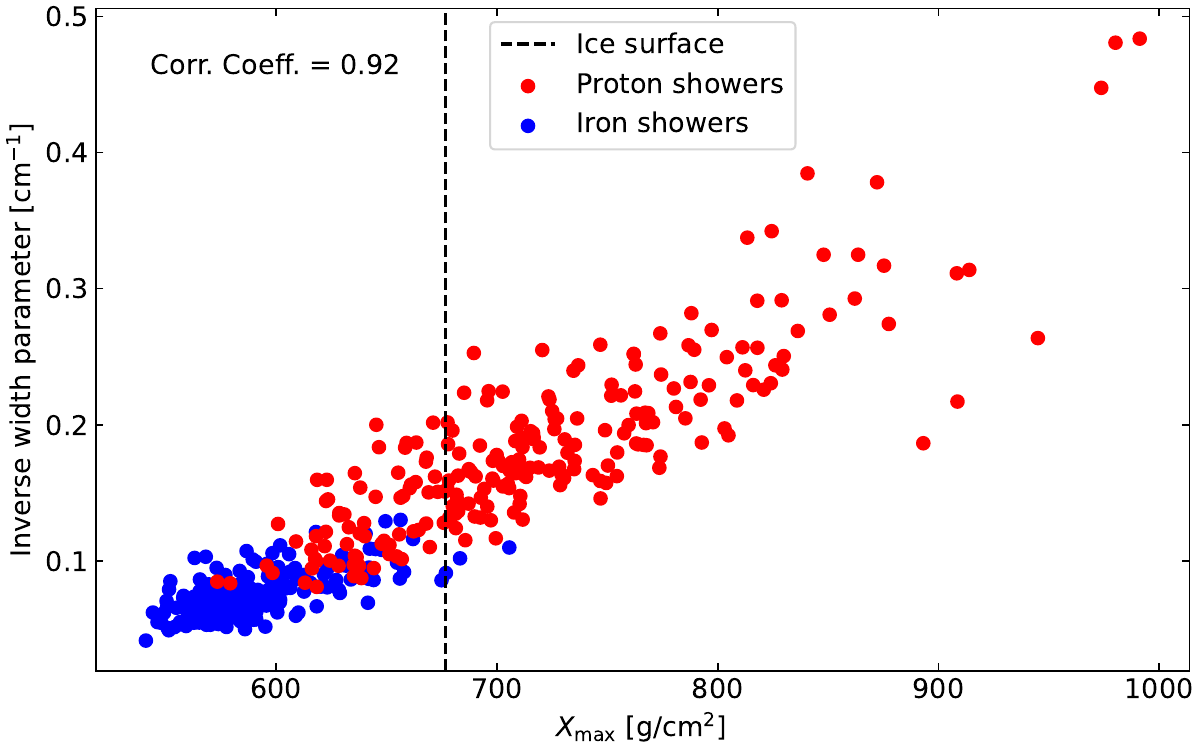}
  \caption{
Correlation of the inverse lateral in-ice core width with the \Xmax observable. 
The low-pass cutoff frequency of the radio spectrum depends on the inverse core width, making the inverse width parameter a proxy for a radio discriminant of the primary CR mass. 
  }
  \label{fig:xmax_correlation}
\end{figure}

\subsection*{Comparison with $X_\mathrm{max}$}
\noindent
Finally, we calculate the correlation of the in-ice core width with the \Xmax observable.
We simulate vertical showers of fixed energy $E=10^{17}$\,eV.
The \Xmax is calculated by fitting a Gaisser-Hillas distribution to the in-air longitudinal shower profile.
The in-ice profile is not included, since the excess particle production beneath the ice surface distorts the distribution.
Fig.~\ref{fig:xmax_correlation} shows the correlation between the inverse width parameter of the in-ice cascade and $\Xmax$,
displaying a strong positive correlation coefficient of 0.92 across both proton and iron showers.
However, the fact that the correlation is not maximal implies that there is additional information about the shower development to be gained from measuring the in-ice core width via the radio spectrum.

\subsection*{Conclusions}
\noindent
In this contribution, we show that currently operating in-ice radio neutrino observatories are theoretically capable of reconstructing cosmic ray primary energy and mass composition using the in-ice Askaryan radio signal.
Because of the long radio attenuation length, in-ice radio detectors have a unique ability to observe the shower core from afar, using the shape of the radio spectrum as a proxy. 
Additionally, the large effective volume for each radio station compensates for the low intrinsic flux of cosmic rays above $10^{17}$\,eV, where the mass composition of cosmic rays is useful for understanding the Galactic to extragalactic transition of cosmic rays.
In future work we plan to provide a more detailed study of radio observables for mass discrimination and
explore the dependence of the radio signal on the high-energy hadronic interaction model used for simulating the air shower.
Eventually, we hope that in-ice radio detectors will be able to perform novel cosmic ray analyses,
which will also inform studies of the relevance of Askaryan emission from cosmic rays as a neutrino background. 
So-called ``double-bump" showers~\cite{De_Henau_2025} could also be interesting to study with this technique.

\subsection*{Acknowledgements}
\noindent This work was supported by NSF Award Number 2411662. The University of Chicago Research Computing Center provided computing support. We thank the attendees of the ARENA conference for many insightful comments.

\bibliographystyle{JHEP}

\bibliography{refs}

\end{document}